\documentstyle[11pt,aaspp]{article}

\begin{document}

\title{Theory of Bipolar Outflows from High-Mass Young Stellar Objects}

\author{Arieh K\"onigl}
\affil{Department of Astronomy and Astrophysics, University of
Chicago, 5640 S. Ellis Ave.,}
\affil{Chicago, Illinois 60637, U.S.A.}
\affil{Electronic mail: arieh@jets.uchicago.edu}




\begin{abstract}
There is a growing number of observational indicators for the
presence of bipolar outflows in massive, young stellar objects
that are still accreting mass as part of their formation process.
In particular, there is evidence that the outflows from these
objects can attain higher velocities and kinetic luminosities
than their lower-mass counterparts. Furthermore, the higher-mass
objects appear to smoothly continue the correlation found in T
Tauri stars between outflow and accretion signatures, and in
several cases there are direct clues to the existence of a
circumstellar disk from optical and infrared imaging and
spectroscopy as well as from millimeter-wavelength
interferometry. These results suggest
that the disk--outflow connection found in low-mass
pre--main-sequence stars extends to more massive objects, and
that a similar physical mechanism may drive the outflows in both
cases. I examine the observational basis for this hypothesis and
consider how the commonly invoked centrifugally driven wind
models of bipolar outflows in low-mass stars would be affected
by the various physical processes (such as photoionization,
photoevaporation, radiation pressure, and stellar wind ram
pressure) that operate in higher-mass stars. I then list some of
the interesting questions that one could hope to address as this
young field of research continues to develop.

\end{abstract}




\section{Introduction}

The subject of accretion-driven outflows from luminous,
high-mass stars is still a fairly new area of
research. In order to define the framework within which the
significance of the various observational findings could be
assessed and meaningful theoretical models formulated,
it is useful to first review what is already known, and what the
remaining open questions are, in the more mature field of
research concerning accretion-driven outflows from
low-luminosity stars.

There are now over 150 catalogued optical outflow
sources associated with low-luminosity ($L_{\rm bol} < 10^3 \,
L_{\sun}$) young stellar objects (YSOs). They appear as
high-velocity (radial speeds $\sim 200-400$ km s$^{-1}$) ionized
and neutral gas {\it jets} and as {\it bipolar}\/ (molecular) {\it
flows}, which evidently represent ambient gas entrained and
driven by the jets (see Edwards et al. 1993a and Bachiller 1996
for reviews). There
is a strong apparent correlation between the presence of {\it outflow}\/
signatures (such as P Cyg line profiles, forbidden line emission,
and thermal radio radiation) and {\it accretion disk}\/
diagnostics (such as ultraviolet, infrared, and millimeter
excess emission) in these sources (e.g., Hartigan et
al. 1995). Most notably, the so-called classical T Tauri stars (cTT's)
consistently exhibit both types of properties, whereas the
weak-lined T Tauri stars (wTT's), which in most other respects
closely resemble cTT's, lack {\it both}\/ outflow {\it and}\/
accretion characteristics. Direct evidence for the presence of
disks in YSOs has been obtained from millimeter and
submillimeter interferometric mappings, which have resolved the
structure and velocity fields of disks down to scales of a few
tens of AU (e.g., Sargent 1996; Guilloteau et al. 1997; Kitamura
et al. 1997; Wilner and Lay 1999). High-resolution images of
disks in low-luminosity YSOs have also been obtained in the
near infrared (NIR) using adaptive optics and in the optical using the
{\em Hubble Space Telescope} (e.g., Stapelfeldt et al. 1997;
McCaughrean et al. 1999).

Another important observational finding in low-mass ($M \la 2 \,
M_{\sun}$) YSOs (where,
from here on, ``low-$M$'' is used interchangeably with ``low-$L_{\rm bol}$'')
is that many of them have been inferred to possess
a strong ($\la 10^3$ G) stellar magnetic field that truncates
the disk at a distance of a few stellar radii from the YSO and channels
the flow toward high-latitude accretion shocks on the stellar surface. 
The evidence for this comes from the detection of periodic surface
``hot spots'' (e.g., Herbst et al. 1994) as well as from
spectral line profiles, particularly of the upper
Balmer lines and Na D (e.g., Edwards et al. 1994), Br$\gamma$ (e.g.,
Najita et al. 1996a), He I and He II (e.g., Guenther and
Hessman 1993; Hamann and Persson 1992; Lamzin 1995), and the Ca
II infrared triplet (e.g., Muzerolle et al. 1998). Direct
measurements of stellar magnetic field strengths are difficult,
but several kilogauss-strength detections have already been
reported (e.g., Basri et al. 1992; Guenther et al. 1999;
Johns-Krull et al. 1999). The magnetic interaction
between the star and the
disk could in principle account for the typically low rotation
rates of cTT's (e.g., K\"onigl 1991) as well as for the
systematically shorter rotation periods measured in wTT's
(Bouvier et al. 1993; Edwards et al. 1993b).

Finally, it is worth mentioning that accretion onto low-mass
YSOs is evidently nonsteady. In particular, these objects
exhibit episodic accretion events that have been inferred to
last $\sim 10^2$ yr and to repeat on a time scale of $\sim 10^3$
yr during the initial $\sim 10^5$ yr of the YSO lifetime (e.g.,
Hartmann and Kenyon 1996). The mass accretion rate during these
episodes is quite high, and it has been estimated that most
of the mass that ends up in the central star could be accreted in
this fashion. It has also been determined that these so-called
FU Orionis outbursts give rise to high-velocity gas outflows
that originate at the surfaces of the circumstellar accretion
disk (Calvet et al. 1993).

The current ``paradigm'' of bipolar outflows in low-$M$ YSOs,
which attempts to interpret the above observational results, can
be summarized as follows. The outflows are powered by accretion,
and probably represent centrifugally driven winds from the disk
surfaces (see K\"onigl and Ruden 1993 and K\"onigl and Pudritz
1999 for reviews). The accretion and
outflow are mediated by a magnetic field that
corresponds either to interstellar field lines that had been advected by the
inflowing matter (e.g., Wardle and K\"onigl 1993) or to a stellar,
dynamo-generated magnetic field (e.g., Shu et al. 1994). The
origin of the field (and, correspondingly, the origin of the
outflow in relation to the YSO), as well as the precise manner
by which a sufficiently strong open field configuration is
maintained along the disk, or, alternatively, the manner by
which a stellar field can both channel an inflow and drive
an outflow, are among the key issues of the theory that are not
yet fully resolved. The currently favored interpretation of FU
Orionis outbursts is that they represent a dwarf nova-like
thermal instability in the innermost, weakly ionized (in
quiescence) region of the disk. The effect of a magnetic field
on the evolution of this instability and its possible role in
driving the associated outflows are other important open
questions in the theory.

Having outlined the relevant observations of low-luminosity
YSOs, I now turn to examine the data on high-luminosity
outflow sources. I focus attention on
pre--main-sequence (PMS) stars and examine, first, whether the observations of
higher-mass YSOs can be interpreted within the same framework as
their lower-mass counterparts, and, second, whether the new
information on high-luminosity outflow sources can shed light on
any of the outstanding questions in the theory of low-$L_{\rm
bol}$ YSOs.

\section{Observations of Outflows from High-Luminosity Stars}

Energetic outflows from luminous young stars have been
detected by similar means to those used in identifying bipolar outflows
in low-luminosity YSOs, namely, through molecular line emission
from the swept-up ambient gas, and through optical and radio
emission from the ionized gas component in stellar
jets. Since high-mass YSOs are often found in regions of
low-mass star formation, confusion with low-luminosity objects
may complicate the determination of the flow structure as well
as the identification of the driving source (which, for example,
may be based on the presence of an isolated IRAS source or of an
ultracompact HII region on or near the flow axis). For example, in the case
of NGC 2024, Chernin (1996) has argued (on the basis of
$4\arcsec$-resolution maps) that several outflows are,
in fact, present in the region and that they do not appear to be driven by the
known far-infrared sources. He suggested that the outflows might
be driven, instead, by as yet unidentified low-mass
stars.\footnote{In this connection it is worth noting that NIR
speckle interferometry of HAeBe stars has revealed that a significant
($31\pm10\%$) fraction of them possess a close IR companion with
a projected separation in the range $50-1300 \ {\rm AU}$ (Leinert
et al. 1997).} In a similar vein, radio continuum sources
interpreted as ultracompact HII regions
could instead trace the sites of shock excitation by the
outflow: such a misidentification has, for example, been claimed
to have occurred in the case of the outflow from the
high-luminosity YSO Cep A (see Corcoran et al. 1993 and
Hughes and Wouterloot 1984). Nevertheless, the
number of bipolar flows studied with adequate resolution by
means of molecular line interferometry has been gradually
increasing, and there is accumulating evidence that, as
in the case of low-mass objects, they are a common
property also of newly formed high-mass stars (see Richer et
al. 1999 for a review). It appears
that the higher-luminosity objects typically produce less
well-collimated molecular flows than their lower-$L_{\rm bol}$
counterparts, although this could possibly be due to the fact
that these outflows tend to emerge from their surrounding
molecular cloud cores at a relatively early stage. The basic spatial
and kinematic structures of the flows do not, however, seem
to depend on the underlying source luminosity, and the
momentum deposition rate in the outflow evidently increases as a
simple power law of the luminosity for $L_{\rm bol}$ ranging all
the way from $\sim $1 to $\sim 10^6 \, L_{\sun}$.

Optical observations of jets from high-luminosity sources are
subject to several detection biases (Mundt and Ray 1994), including short
evolutionary timescales, typical association with comparatively
distant star-formation regions, and confusion by bright,
extended reflection nebulae as well as by background HII
emission. Another bias can be traced to the effect of the
ionizing flux from the central object (see Fig. 1
below). Despite these complicating
factors, a significant number of sources with outflow signatures
have already been detected. Mundt and Ray (1994) list 24 examples of
optical jets associated with Herbig Ae/Be stars (HAeBe's) and
other high-luminosity YSOs, whereas Corcoran and Ray (1997)
report that 28 out of 56 HAeBe's that they studied had
detectable [OI]$\lambda$6300 forbidden line emission. As
discussed in the above references, the jet outflow speeds in high-luminosity
($L_{\rm bol} \ga 10^3 \, L_{\sun}$) sources lie in the range
$\sim 600-900$ km s$^{-1}$, which are a factor $\sim 2-3$ higher
than the corresponding speeds in the low-$L_{\rm bol}$ sources,
and have inferred mass outflow rates that are a factor $\sim
10-100$ times higher than in the low-luminosity YSOs. There is
also an indication that a larger fraction of the jets in luminous
sources are poorly collimated.

\section{The Accretion Disk Connection}

There is now growing evidence that the correlation found in
low-mass YSOs between the signatures of energetic outflows and
accretion disks extends also to the more massive HAeBe stars.
Corcoran and Ray (1997) discovered that, in most cases, the centroid
velocity of the low-velocity component of the
[OI]$\lambda$6300 emission line is {\it blueshifted}\/ with respect
to the stellar rest velocity. The same behavior is found in cTT's and has
been convincingly interpreted as evidence for the presence of
extended, optically thick disks that block the
redshifted line-emission region from our view. The forbidden emission lines in
cTT's often exhibit both a low-velocity component (LVC), which has been
attributed to a disk-driven outflow, and a high-velocity
component (HVC), whose interpretation is still controversial but
which evidently originates in the vicinity of the YSO.
The HVC is also observed in the [OI] line emission
from some HAeBe's, but it is found less frequently than in
cTT's (see also B\"ohm and Catala 1994). The latter finding
was attributed by Corcoran and Ray (1997) to an
evolutionary effect (wherein the HVC disappears before the LVC
as the outflow activity gradually diminishes), although it is
conceivable that at least in some luminous sources the absence
of a high-velocity neutral oxygen component may be the result of
photoionization near the outflow axis (S. Martin
1994, personal communication). In view of the fact that the
ionization potential of neutral oxygen is nearly identical to
that of hydrogen, one would not expect to detect [OI] emission
within the Str\"omgren surface bounding the HII region around
the star. If the disk is a source of a centrifugally driven
outflow, the density distribution around the star will be highly
stratified (e.g., Safier 1993) and the Str\"omgren surface will have a
roughly conical shape centered on the symmetry axis (see
Fig. 1). Under these circumstances, the HVC [OI] emission, produced
near the stellar surface, will be absent, but the LVC,
presumably generated above the disk surface further out in a
region that is shielded from the ionizing radiation, will be detectable.
This interpretation is supported by observations of a source
like LkH$\alpha$ 234, in which a well collimated, high-velocity
jet is detected (Ray et al. 1990)\footnote{It is has been
suggested, however, that the jet in this
source is driven by a cold mid-infrared companion rather than by LkH$\alpha$234
itself (Cabrit et al. 1997).} even though only a low-velocity [OI]
component is seen in the vicinity of the central star.
\begin{figure}[p]
\plotone{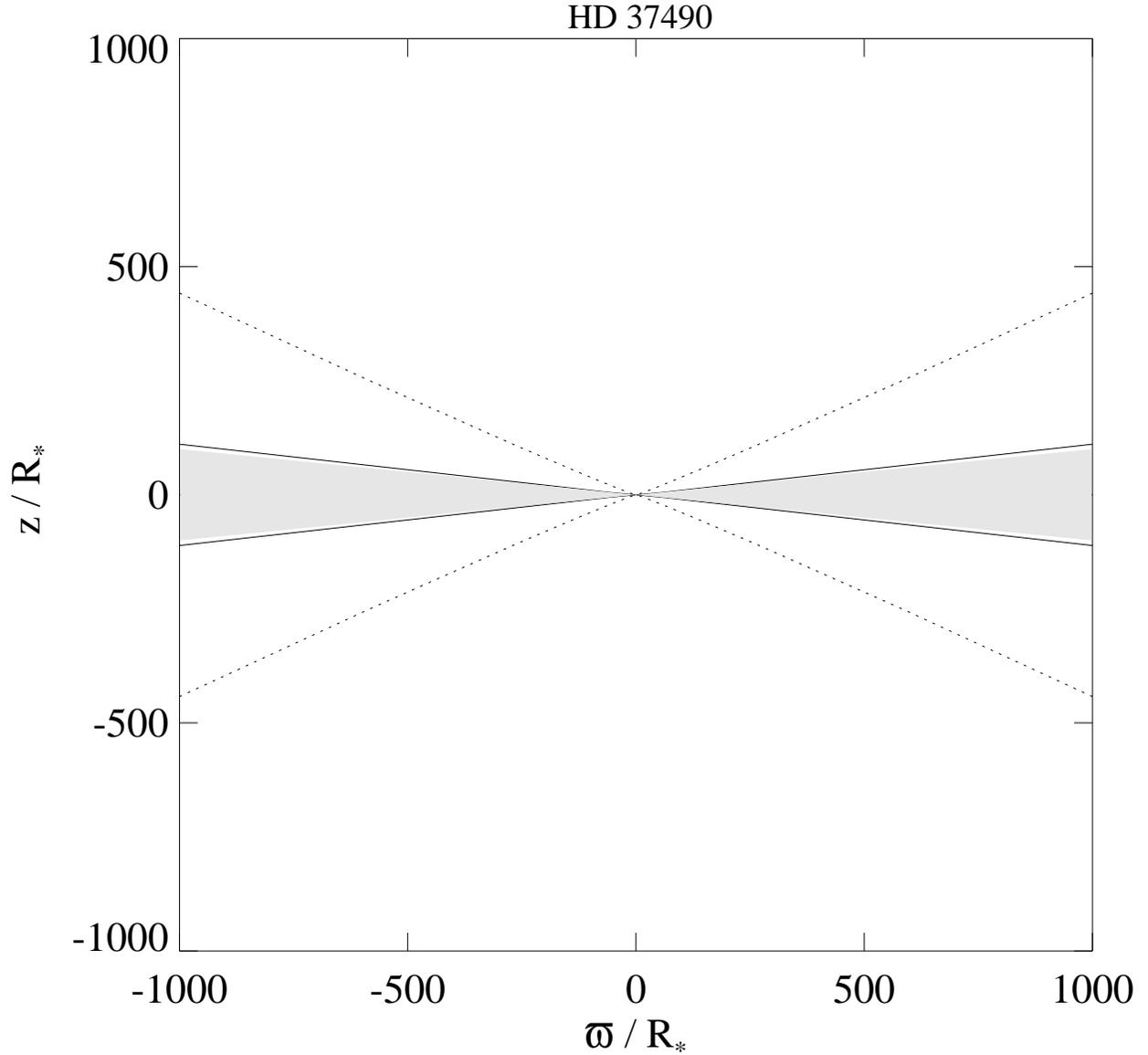}
\caption{Str\"omgren surfaces in a disk-driven wind near a
high-luminosity YSO, calculated using the direct stellar
ionizing flux but neglecting the diffuse radiation field. The
{\it dashed}\/ and {\it solid}\/ lines
represent the shape of the surfaces for mass
outflow rates (in the cylindrical radius range $\varpi = 0.1 - 1
\, {\rm AU}$) of $10^{-6} \, M_{\sun} \,
{\rm yr}^{-1}$ and $10^{-7} \, M_{\sun} \, {\rm yr}^{-1}$,
respectively, assuming wind model B of Safier (1993) and the
parameters of the HAeBe star HD 37490 ($L_{\rm bol} = 2
\times 10^4 \, L_{\sun}$, $T_{\rm eff} = 20,400$ K, $R_* = 11.4
\, R_{\sun}$). The shaded region represents an accretion disk of
height $h(\varpi) = 0.1\varpi$. (Courtesy of S. Martin.)}
\label{fig-1}
\end{figure}

Another robust correlation, identified by Corcoran and Ray (1998),
relates the [OI]$\lambda$6300 line luminosity (a signature of an outflow) and
the infrared excess luminosity (a possible signature of a
disk). It appears that the relationship between these two
quantities, originally found in cTT's (e.g., Cabrit et
al. 1990), extends smoothly to YSOs with masses of up to $\sim 10\, M_{\sun}$
and spans 5 orders of magnitude in infrared luminosity. Corcoran
and Ray (1998) analyzed additional correlations between the
forbidden-line and NIR emission properties of HAeBe's
and pointed out that they all follow the same trends as in
cTT's. They also found that all the HAeBe stars in
their sample that exhibit both forbidden line
emission and IR excesses have NIR colors that are consistent
with the presence of an optically thick disk or a disk surrounded by
a dusty envelope. Previous infrared studies of HAeBe's
(e.g., Hillenbrand et al. 1992) have revealed
that many of these objects show infrared excesses with a spectral shape
$\lambda F_{\lambda} \propto \lambda^{-4/3}$ ($\lambda \ga 2.2
\, \mu$m). Such spectra are characteristic of optically
thick disks that are either ``active'' viscous accretion disks
or ``passive'' reprocessing flat disks. The apparent spectral
decline below $\sim 2.2 \, \mu$m has been interpreted as indicating the
presence of effective ``holes'' in the optically thick disks on
scales $\sim 3-25$ times the stellar radius (see also Lada and
Adams 1992). It was originally
suggested that the holes could represent regions where the disk
is either truncated by a stellar magnetic field or else is optically thin. 
For reasonable accretion rates, a stellar magnetic field is
unlikely to truncate a disk beyond a few stellar radii (e.g.,
K\"onigl 1991). However, the innermost regions might be
optically thin because of a low local mass accretion rate (Bell
1994; see \S 5). The mass accretion rate required to reproduce
the $\sim 3 \, \mu$m peak in the NIR spectral energy
distribution is too large ($\ga 10^{-6} \, M_{\sun} \, {\rm
yr}^{-1}$) for the innermost disk regions to remain optically
thin, but Hartmann et al. (1993) argued that the observed peak
could, instead, be due to the transient heating of grains
in a dusty envelope by ultraviolet photons from the central star
(see also Natta et al. 1992, 1993).

The interpretation of the infrared and
sub-mm spectra of HAeBe's in terms of disks has not been
universally accepted: several authors have, in fact, claimed
that the spectra can be explained entirely in terms of dusty
spherical envelopes (e.g.,
Miroshnichenko et al. 1997; Pezzuto et al. 1997). It was similarly suggested
that much of the millimeter emission in these systems arises
in extended envelopes, and, furthermore, that the contribution
from ionized gas may have led to an overestimate of the dust
emission in many sources (e.g., Di Francesco et al. 1997). 
Furthermore, in some cases there are indications that the measured
far-infrared emission may not even arise in the immediate
vicinity of the HAeBe's (Di Francesco et al. 1998).
However, several strong disk candidates have by now been
identified by mm-wavelength interferometry (e.g., Mannings and Sargent
1997). Among the sources observed by Mannings and Sargent, two
appear as elongated molecular line-emission regions and exhibit
ordered velocity gradients along their major axes, which is strongly
suggestive of the presence of rotating disks. The disk radii and
masses determined by
these authors are similar to those found in cTT's,
although, in view of the short clearing
time of optically thick disks inferred for HAeBe's ($\sim 0.3\,
{\rm Myr}$, as compared with $\sim 0.3\, {\rm Myr}$ for cTT's;
Hillenbrand et al. 1992), this may reflect the comparatively large age of the
objects in their sample ($\sim 5-10\, {\rm Myr}$, compared to $\sim
1\, {\rm Myr}$ for typical cTT's). Further support for the
presence of disks around HAeBe's has come from adaptive-optics
IR imaging polarimetry and {\it HST}\/ optical
images of the object at the origin of the R Mon
outflow, which were interpreted in terms of a $\sim 10^2\ AU$ optically thick
accretion disk surrounding a $\sim 10\, M_{\sun}$ HAeBe star
(Close et al. 1997). 

Another suggestive piece of
evidence is the detection in several HAeBe's of CO overtone bandhead
emission that exhibits broadening by a peristellar
velocity distribution that scales with radius as
$v(r) \propto r^{-1/2}$ (e.g., Chandler et al. 1995; Najita et
al. 1996b). This velocity field is consistent with Keplerian
rotation and the emission has therefore been attributed to a
circumstellar disk. An alternative interpretation
(which, like the disk model, also applies to
low-luminosity YSOs in which CO bandhead emission has been detected)
is that the emission originates in a magnetic accretion funnel
that channels the inflowing matter from an accretion disk,
with the observed broadening produced as the gas free-falls along the
stellar magnetic field lines toward the stellar surface (Martin
1997). There are, in fact, other
tantalizing observational clues that point to the presence of magnetospheric
accretion in certain HAeBe's. These include, in particular, the
detection of inverse P Cygni (IPC) H$\beta$ line profiles in a number
of such stars (see Fig. 2). In one survey of HAeBe
stars (Ghandour et al. 1994), 4 out of 29
objects were found to show clear evidence for such profiles, with the
redshifted absorption feature occurring between $\sim 100$ and $\sim
700 \, {\rm km} \, {\rm s}^{-1}$ relative to the rest
velocity (L. Hillenbrand 1996, personal
communication).\footnote{For comparison with
this $\sim 14 \, \%$ apparent frequency, $40 \, \%$ (6/15) of the cTT's
surveyed by Edwards et al. (1994) exhibited IPC
profiles in H$\beta$. The detection frequency for the above
sample of HAeBe's would, however, increase to $\sim 25 \,
\%$ if one also included objects with
a more tentative IPC classification; see Ghandour et al. 1994.}  Sorelli et
al. (1996) proposed a similar
interpretation of the redshifted Na D absorption components observed in
several HAeBe's. As the latter authors have noted, the
Na D lines are likely to originate in a region containing
neutral hydrogen, which could give rise to detectable Ly$\alpha$
absorption features. These may, however, prove difficult
to identify in the presence of broad emission features
originating in an associated outflow.\footnote{Blondel et al. (1993)
detected Ly$\alpha$ emission lines in several HAeBe stars and
attributed them to radiation from magnetic accretion funnels;
however, one can argue that a wind origin is more likely in this
case (L. Hartmann 1996, personal communication).} It is, however,
important to bear in mind that alternative
interpretations of the IPC profiles have been proposed. For example,
they have been attributed to the evaporation of comet-like
bodies that reach the vicinity of the star (e.g., Grinin et
al. 1994) as well as to ongoing quasi-spherical infall onto the YSO
(Bertout et al. 1996). A detailed comparison between the data and
the specific predictions of each model would therefore be necessary
before one could accept the presence of IPC profiles in these
stars as unequivocal evidence for magnetospheric
accretion (see Sorelli et al. 1996). Firm evidence might
be provided, for example, by the discovery of periodic surface ``hot spots''
of the type previously found in some cTT's,
although the low expected contrast between the effective temperatures of
the accretion shock and of the surrounding stellar photosphere in a
hot star would make such detections difficult.
\begin{figure}[p]
\plotone{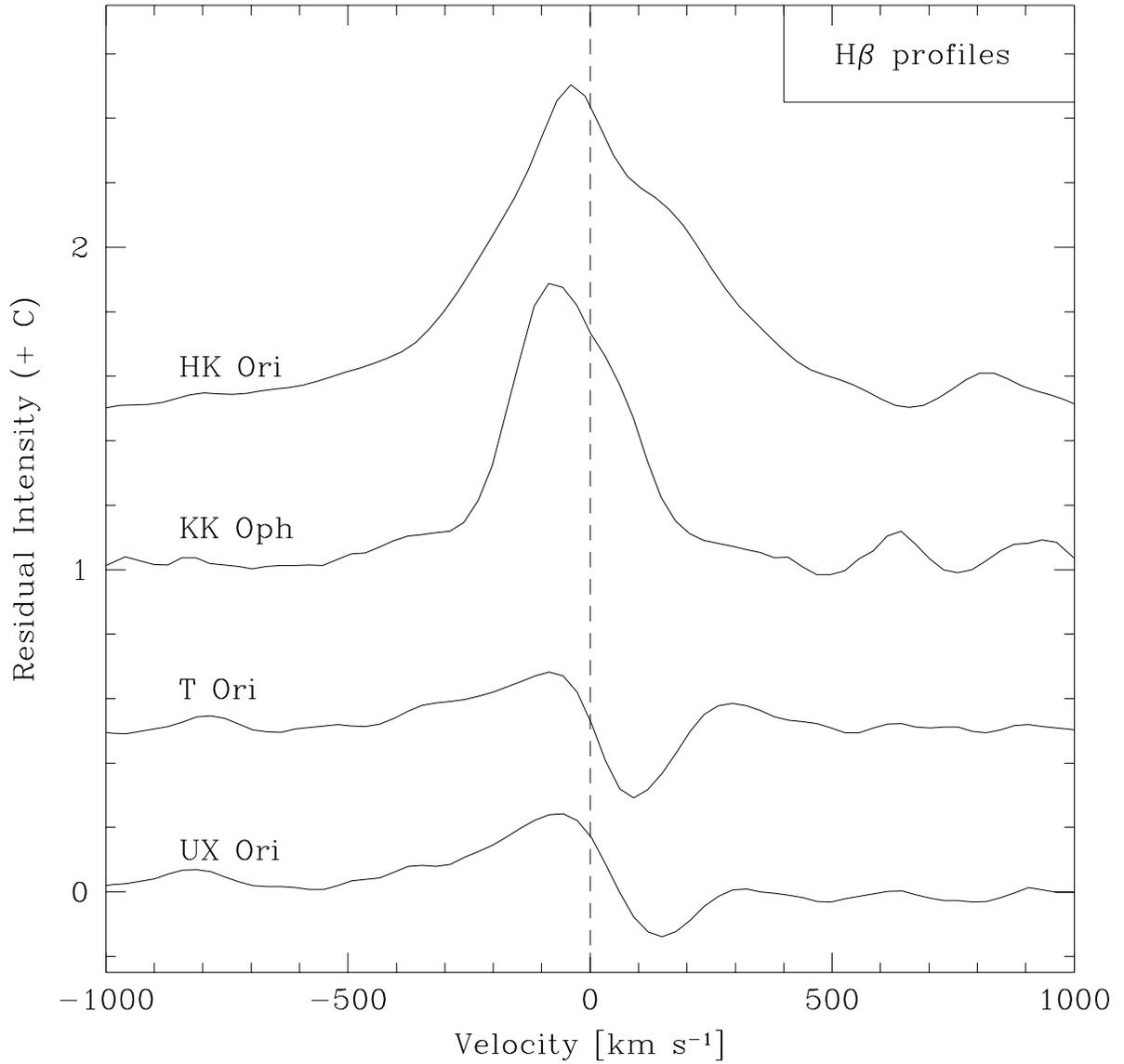}
\caption{Residual H$\beta$ spectra of HAeBe stars
with IPC profiles, obtained by subtracting in each case a
standard photospheric spectrum for a star of the same spectral
type. The displayed spectra are vertically offset from each
other for the sake of clarity and have normalized intensities. These are the
best examples of IPC profiles among a sample of 29 HAeBe's
observed in 1992 at the KPNO 2.1m telescope. (Courtesy of L. Hillenbrand.)}
\label{fig-2}
\end{figure}

Disk signatures (including NIR excess, optical veiling, and CO
bandhead emission) have been reported also in higher-mass
($\sim 5-20\, M_{\sun}$) stars (Hanson et al. 1997). These findings
are of particular interest, since, for a representative mass
inflow rate $\dot M_{\rm in} = 10^{-5} \, M_{\sun} \, {\rm
yr}^{-1}$, stars with masses $M > 8 \, M_{\sun}$ are predicted
to reach the main sequence before disk accretion
has ceased (Palla and Stahler 1993). Such disks may therefore be
expected to show the effects of the interaction with the strong
radiation field and stellar wind that are produced by a high-mass
main-sequence star. The direct imaging of disk candidates
in high-$M$ stars and a comparison between their properties and those of
disks in low-$M$ YSOs are thus an important challenge for
future observations. Since the structure
of a circumstellar disk likely depends on how the
central star is formed, these studies could also help to test the
suggestion (Stahler et al. 1999) that stars with masses $M \ga 10 \,
M_{\sun}$ are assembled through the coalescence of lower-mass objects
at the centers of young stellar clusters. One piece of evidence
that has been cited in support of this interpretation is the apparent
stellar clustering around HAeBe stars: Testi et
al. (1997; see also Hillenbrand 1995) found that such clustering becomes
significant for stars of spectral type B7 or earlier and
argued that this is consistent with intermediate-mass YSOs
representing a transition between the low-mass and high-mass modes
of star formation. If so, then the study of disks in HAeBe's
could also be useful for testing the coalescence picture.

It is noteworthy that several independent studies have
established correlations of the type $\dot
M \propto L_{\rm bol}^{0.6}$ for the mass {\it accretion}\/ rate
(from IR continuum measurements; Hillenbrand et al. 1992), the
{\it ionized}\/ mass {\it outflow}\/ rate in the jets (from radio continuum
observations; Skinner et al. 1993), and the bipolar {\it
molecular outflow}\/ rate (from CO line measurements; Levreault
1988) in both low-luminosity and high-luminosity YSOs. Taken
together, these relationships suggest that a strong
link between accretion and outflow exists in both low-mass and
high-mass stars.

\section{Modeling Issues}

The detection of similar accretion and outflow signatures in
low-$L_{\rm bol}$ and high-$L_{\rm bol}$ YSOs and the evidence for a
strong correlation between them that continues smoothly from
low- to high-luminosity objects provide strong arguments in
favor of a similar underlying physical mechanism operating in all newly
formed stars. In particular, the disk-driven hydromagnetic wind
scenario, which is currently the leading model for the origin of
bipolar outflows in low-$L_{\rm bol}$ stars (see \S 1), may also
apply to HAeBe's and even higher-mass stars. It is important to
note, however, that the basic model worked out for the
low-luminosity objects would need to be extended and modified by
the inclusion of several new effects that are specific to
high-luminosity objects. Some of these effects have already been
considered before in a different context, but incorporating them
all together into a self-consistent accretion/outflow model is
one of the main theoretical challenges in this new field of research.
Among the anticipated new elements of the theory, one can list
the following:
\begin{itemize}
\item Enhanced field--matter coupling near the disk surface due
to both the direct and the diffuse ionizing radiation from the central star,
leading to higher mass accretion and outflow rates (e.g.,
Pudritz 1985).
\item Disk photoevaporation (e.g., Hollenbach et
al. 1994; Yorke and Welz 1996), creating a low-velocity ($v$ of
order the sound speed $c_{\rm s} \approx 10 \,
{\rm km} \, {\rm s}^{-1}$) disk outflow beyond the
``gravitational radius'' $r_{\rm g} = GM/c_{\rm s}^2 \approx
10^{15}(M/10 \, M_{\sun}) \, {\rm cm}$ (or even further in the
presence of a strong stellar wind). Photoevaporation may
facilitate the injection of mass
into a centrifugally driven disk outflow, but a hydromagnetic
wind from the inner disk could reduce the mass evaporation rate
further out by intercepting some of the ionizing radiation
from the central star.
\item A strong, radiatively driven {\em stellar} wind, which may be transformed
into a highly collimated jet through a dynamical interaction
with a disk-driven wind (e.g., Frank and Mellema 1996) or a disk
magnetic field (e.g., Kwan and Tademaru 1995). A strong,
radiatively driven outflow may also be induced in the {\em disk}
by the intercepted stellar radiation: this outflow is predicted
to originate within a few stellar radii from the YSO's surface,
to be predominantly equatorial, and to have significantly lower
speeds than those of the stellar wind (Drew et al. 1998).
\item Radiation pressure effects on dust. Because of the large
dust scattering cross section at UV and optical wavelengths, the
effective Eddington luminosity for a dusty gas is much lower
than the electron-scattering critical luminosity, and is given
by $L_{\rm crit, \, dust} \approx 4 \times 10^2 (M/10
M_{\sun}) \, L_{\sun}$ (e.g., Wolfire and Cassinelli
1987). Radiation pressure effects could thus be important beyond
the dust sublimation radius $r_{\rm sub} \approx 1 \, (L_{\rm
bol}/10^2 \, L_{\sun})^{1/2} \, {\rm AU}$ and might contribute
to the flow acceleration and also lead to the ``opening up'' of
the streamlines (see K\"onigl and Kartje 1994). The latter effect
could be at least partially responsible for the apparently lower degree of
collimation of jets from high-$L_{\rm bol}$ sources (Mundt and Ray
1994; see \S 2).
\item Photoionization heating and radiative excitation. The
strong stellar radiation field is expected to be the main
heating mechanism of the gas in the stellar vicinity. In
particular, it may dominate the ambipolar diffusion heating that
is important for weakly ionized outflows in low-luminosity YSOs
(Safier 1993) and could, in fact, cut it off altogether within
the Str\"omgren surface. Radiative excitation may give rise to
unique emission signatures, including Ly$\alpha$ lines (e.g.,
Blondel et al. 1993) and enhanced overtone emission from the
higher CO bandheads (Martin 1997).
\end{itemize}

The incorporation of low-luminosity and high-luminosity YSOs into the
same theoretical framework may help resolve some
of the outstanding issues in the modeling of bipolar outflows
from low-mass stars. For example, the origin of the LVC and HVC
forbidden line emission is not yet fully
understood. Magnetically driven outflows have been leading
candidates for their interpretation, but both stellar
field-based (e.g., Shang et al. 1998) and disk field-based (e.g.,
Cabrit et al. 1999) models have been proposed. As noted
in \S 3, the apparent decrease in the detection frequency of
the [OI] HVC (relative to the [OI] LVC) in higher-luminosity YSOs might
be related to the locations of the HVC and LVC emission regions with respect
to the star. If so, this could prove useful in the attempt to
discriminate between the competing models.

\section{Further Questions}

\subsection{The Role of a Stellar Magnetic Field in Channeling
Accretion and Driving an Outflow}

In one of the proposed models for the origin of bipolar outflows and
jets from low-mass stars, the stellar magnetic field plays a pivotal role
in the generation of the underlying centrifugally driven wind
(e.g., Shu et al. 1999). If all YSO outflow sources can indeed
be described by the same general model (see \S 4), then this
interpretation requires that HAeBe's (as well as higher-mass stars)
should have a strong stellar magnetic field if they give rise to
an energetic outflow.\footnote{As discussed by Catala
et al. (1986), at least $\sim 20\%$ of HAeBe's (those belonging to
the P Cygni subclass; see Finkenzeller and Mundt 1984) give rise
to powerful winds with mass loss rates in the range $\sim
10^{-8}-10^{-6} \, M_{\sun} \, {\rm yr}^{-1}$. The existence of a stellar
magnetic field is also required in models that attribute the
acceleration of these winds to hydromagnetic waves (e.g.,
Strafella et al. 1998).} As was pointed
out in \S 3, there is, in fact, persuasive evidence in at least
some HAeBe's for the existence of a magnetic field that is
strong enough to channel accreting gas onto the stellar
surface. There is also evidence for a strong,
ordered magnetic field in {\it non}\/accreting HAeBe
stars. In particular, VLBI radio measurements have revealed the presence of
extended ($\ga 3-10 \, R_*$), organized magnetic field
configurations that are similar to those observed in certain
wTT's, and which also resemble the field structures of magnetic
Ap and Bp stars, in at least two such objects (Andr\'e et al. 1992). 

The evidence for strong magnetic fields in at least some HAeBe's
is puzzling, since such stars\footnote{HAeBe stars are
commonly taken to be YSOs in the mass range $2 \, M_{\sun}
\la M \la 8 \, M_{\sun}$.} are not
expected to have a deep convective layer, and therefore,
according to the conventional picture, could not generate a
strong field through dynamo action. In fact, according to the
calculations of Palla and Stahler (1993), stars that accrete at a
constant rate of $10^{-5} \, M_{\sun} \, {\rm yr}^{-1}$ during
the PMS phase should be fully convective for $M
< 2.4 \, M_{\sun}$, have a subsurface convection layer due to deuterium
burning for $2.4 \, M_{\sun} < M < 3.9 \, M_{\sun}$, and be
fully radiative for $M > 3.9 \, M_{\sun}$. One possible
resolution of this apparent difficulty is that the observed
structures represent a fossil magnetic field and that HAeBe's
possessing such strong fields are, in fact, the precursors
of the main-sequence Ap/Bp stars (Andr\'e et al. 1992). The
association with Ap/Bp stars could in principle be tested by
comparing the rotation rates of HAeBe's that are strong nonthermal
radio sources (accepting Andr\'e et al.'s argument that such
emission is a reliable tracer of a large-scale, organized stellar field) with
the rotation rates of other HAeBe's. As a class, HAeBe's
are intermediate rotators (rotating at $\sim 0.3$ of breakup
with mean projected speeds of $\sim 100 \, {\rm km} \, {\rm
s}^{-1}$; e.g., B\"ohm and Catala 1995), but those objects that
have extended magnetospheres may be expected to rotate more
slowly, in accordance with the observed trend in Ap/Bp stars. It
remains to be explained, however, how a closed magnetic field
configuration would arise from the unipolar field that is
likely to be incorporated into the star during the PMS accretion
phase (e.g., Li and Shu 1996).

An alternative interpretation of the field is that it originates
in a stellar dynamo that taps directly into the
large rotational energy reservoir of the star without requiring
convection to also be present. Tout and Pringle (1995) and
Ligni\`eres et al. (1996) have explored specific models
along these lines. A dynamo mechanism is a promising candidate
for the enhanced surface activity exhibited by HAeBe stars
of the P Cygni subclass (e.g., B\"ohm and Catala 1995), and it
may also account for the X-ray emission detected toward
a fair number of HAeBe's (Zinnecker and
Preibisch 1994).\footnote{Zinnecker and Preibisch (1994)
discuss stellar wind shocks as the most likely source of the
X-rays. However, Skinner and Yamauchi (1996), analyzing ASCA data
for a bright Herbig Ae star, have concluded that the X-rays in
that case originate from the immediate stellar vicinity,
although possibly from a late-type companion. In fact, confusion
with nearby low-mass YSOs is a potentially serious problem for
measurements with existing X-ray telescopes, whose angular
resolution typically does not exceed $\sim 5\arcsec$.} Tout and Pringle's
(1995) model requires the mean poloidal field to be rather weak
and thus cannot account for either the large-scale organized field
or the magnetically channeled accretion inferred in
several of these objects. It is also unclear how any model that
relates the dynamo action to the stellar rotation can explain
the apparent lack of a correlation between the various activity
tracers and the projected rotation speed (Zinnecker and Preibisch
1994; B\"ohm and Catala 1995), which contrasts with
the observed trend in T Tauri stars (e.g., Neuh\"auser
1997).\footnote{Note, however, that if some of the observed
activity in HAeBe's is powered by magnetically channeled
accretion, then the absence of a clear correlation with the
stellar rotation might be due to the interaction between the
stellar magnetic field and a circumstellar disk, which would tend to reduce the
rotation rate (see \S 1). As a test of the latter effect, it
would be useful to search for a correlation (similar to the one
found in T Tauri stars) between the rotation speeds of HAeBe's
and the strength of their accretion signatures.}
It is interesting to note
in this connection that even the youngest low-mass YSOs, which
likely are not yet fully convective
objects, already show evidence for strong outflows. In fact, the momentum
discharges inferred in these so-called Class 0 sources are a
factor of $\sim 10$ higher than the values
implied by the tight correlation between the momentum discharge
and the source bolometric luminosity that is found in older
(Class I) low-mass YSO's (Bontemps et al. 1996). If the Class 0
outflows are driven by a stellar magnetic field, then
the mechanisms invoked for producing strong
fields in non-convective YSOs could be relevant not
only to HAeBe's but also to very young low-mass objects.
In order to properly address the above questions, it would be
necessary to carry out a self-consistent calculation of the
stellar and magnetic field evolution that would take into
account the the spatial distribution of radiative and convective regions within
the star as well as the effects of field-mediated accretion onto
the YSO, magnetic braking of the stellar rotation, etc.

\subsection{Does the FU Orionis Phenomenon Have an Analog in
Higher-Mass Stars?}

Bell (1994) suggested that the large apparent ``holes'' that have
been inferred from the infrared spectral modeling of HAeBe's
(see \S 3) may be associated with the ``low'' phase of a
thermal ionization instability that operates in the inner disk
region. As was mentioned in \S 1, the ``high'' phase of this
instability has been proposed as the
origin of the FU Ori outburst phenomenon in low-mass YSOs (e.g.,
Bell and Lin 1994). Given that several other arguments also point
to a comparatively low ($\la 10^{-7} \, M_{\sun} \, {\rm
yr}^{-1}$; e.g., Hartmann et al. 1993) current mass accretion rate onto
the central object in a number of HAeBe's, it is conceivable
that a similar mode of nonsteady
accretion is present in higher-mass stars. One possible
check on this idea might be a search for multiple bow shocks
along the associated jets: such shocks have been detected in several
low-$L_{\rm bol}$ objects, and it has been argued that
they are likely associated with the strong disk-outflow episodes
that characterize FU Ori outbursts (e.g., Reipurth 1989). In the
case of the higher-luminosity YSOs it would, however, be
necessary to examine the effect of the strong stellar radiation
field on the evolution of the instability. In particular, one
would need to check to what extent the inner disk could be
maintained in a state of low temperature and ionization during the ``low''
phase of the instability (see van Paradijs 1996 and King et al. 1996).

\subsection{Disk Evolution around Massive Stars}

Accretion disks around high-mass stars are probably the
predecessors of the Vega-type systems first discovered by IRAS; in
particular, $\beta$ Pictoris has likely evolved from an HAeBe
star. The evolution of such disks could in principle be traced
by comparing the high-$M$ analogs of cTT's with the
corresponding analogs of wTT's as well as with Vega-type systems
(e.g., Strom et al. 1991). The analogs of cTT's and wTT's can be identified
through their accretion signatures (in particular, their
infrared spectra; Lada and Adams 1992, Hillenbrand et al. 1992)
as well as through their outflow signatures (in particular,
their forbidden line emission and the H$\alpha$ equivalent width;
Corcoran and Ray 1998). A systematic search for the high-mass
analogs of wTT's may potentially be carried out by means of an
X-ray survey (cf. Casanova et al. 1995), although this
approach is subject to the caveats that the X-ray emission from
HAeBe's might be associated with outflows rather than being
intrinsic to the YSOs (Zinnecker and Preibisch 1994) or that it
originates in close, low-mass companions (e.g., Skinner and Yamauchi 1996). It
is, however, also conceivable that a sufficiently large data base might be
obtained as part of a more general mapping project, such
as the Sloan Digital Sky Survey.

Hillenbrand et al. (1992) concluded from an analysis of a sample
of 47 HAeBe's that the shorter evolutionary timescales of
more massive stars are reflected in the clearing timescales of
their respective disks, with optically thick disks around such
stars surviving for less than $0.3 \, {\rm Myr}$ (as compared
with $\sim 3 \, {\rm Myr}$ for a typical T Tauri star). It has,
in fact, been surmised that the disk clearing time may in some
cases be even shorter than the stellar evolution
time, resulting in a possibly significant shortening
of the evolutionary phase over which these YSOs would
be classified as HAeBe stars (de Winter et al. 1997).  The
faster disk evolution in the more luminous YSOs may reflect the
effects of photoevaporation and strong stellar winds in such
objects (see \S 4), although it is also possible that the
time-averaged mass accretion rate through these disks is higher
(perhaps as a result of a stronger ionizing flux from the
central star; see \S 4). The disk clearing mechanism in YSOs
is still an open question, and its resolution could benefit from
continued comparative studies of low-mass and high-mass
objects. Such studies, coupled with searches for low-luminosity
companions, could also shed light on the issue of
planet formation in protostellar disks (e.g., by
constraining the timescale of planetesimal growth; see Strom et al. 1991).

\acknowledgments

I am grateful to Steve Martin for his valuable input into the
contents of this paper. I also thank Lee
Hartmann, Lynne Hillenbrand, Debra Shepherd and Frank Shu for useful
discussions and correspondence, Lynne
Hillenbrand for kindly providing data in advance
of publication, and Joseph Cassinelli for inspiring this
undertaking. This work was supported in part by NASA grant NAG 5-3687.

\end{document}